\documentclass[conference]{IEEEtran}
\IEEEoverridecommandlockouts
\usepackage{booktabs} 

\usepackage[ruled]{algorithm2e} 
\usepackage{balance}
\usepackage{varwidth}
\usepackage{subfigure}
\usepackage{caption}
\usepackage{cite}
\usepackage{amsmath,amssymb,amsfonts}
\usepackage{algorithmic}
\usepackage{graphicx}
\usepackage{textcomp}
\usepackage{xcolor}
\usepackage{url}

\usepackage[finalnew]{LatexPackage/trackchanges} 
\addeditor{Ander}
\addeditor{DG}

\usepackage{enumitem}
\setitemize{noitemsep,topsep=1pt,parsep=1pt,partopsep=1pt,leftmargin=*}
\usepackage{amssymb}

\def\BibTeX{{\rm B\kern-.05em{\sc i\kern-.025em b}\kern-.08em
    T\kern-.1667em\lower.7ex\hbox{E}\kern-.125emX}}

\begin{document}

\title{Research in Visible Light Communication Systems with OpenVLC1.3 \\
\thanks{The project that gave rise to these results received the support of a fellowship from "la Caixa" Foundation (ID 100010434). The fellowship code is LCF/BQ/ES16/11570019.}
}
\iftrue
\author{\IEEEauthorblockN{Ander Galisteo}
\IEEEauthorblockA{\textit{IMDEA Networks Institute  \&} \\
	\textit{Universidad Carlos III de Madrid}\\
Madrid, Spain \\
ander.galisteo@imdea.org}
\and
\IEEEauthorblockN{Diego Juara}
\IEEEauthorblockA{\textit{IMDEA Networks Institute} \\
	Madrid, Spain \\
	diego.juara@imdea.org}
\and
\IEEEauthorblockN{Domenico Giustiniano}
\IEEEauthorblockA{\textit{IMDEA Networks Institute} \\
Madrid, Spain \\
domenico.giustiniano@imdea.org}

}
\fi
\IEEEoverridecommandlockouts
\IEEEpubid{\makebox[\columnwidth]{978-1-5386-4980-0/19/\$31.00 ©2019 IEEE}
\hspace{\columnsep}\makebox[\columnwidth]{ }}

\maketitle

\begin{abstract}
In this paper, we present the design and implementation of our latest OpenVLC1.3 platform to perform research in Visible Light
Communication Systems. We retain the advantages of the previous versions such as TCP/IP layers support, software programmability and low-cost front-end. We\change[Ander]{increase the throughput at the transport layer to 400 kb/s}{ re-design the transceiver to support higher modulation rates and sensitivity. This allows us to reach a throughput of 400 kb/s}  (a factor of 4 with respect to the 
previous version) and increase the distance by a factor of 3.5. We further improve the software robustness of the system and reduce the 
\change[Ander]{size}{form factor} at similar hardware cost. 
\end{abstract}

\section{Introduction}

Visible Light Communication is gaining significant interest as a medium to connect to the 
Internet~\cite{pureLiFi,Zhang2015,Liu_2011}. In the last few years, a range of applications have been developed with low-end 
Visible 
Light Communication (VLC) platforms: human sensing~\cite{li2015human}, communication with toys~\cite{tippenhauer2012toys}, 
mobile interaction~\cite{zhang2015extending}, indoor localization~\cite{kuo2014luxapose,zhang2016litell} and passive 
VLC~\cite{wang2016passive,xu2017passivevlc}. Industry interest is also resulting in the establishment of the 
IEEE 802.11bb task group, where the objective is to amend the Medium Access Control (MAC) and Physical Layer (PHY) of IEEE 802.11 
with Light Communications~\cite{802_11bb}. 

To solve the lack of an open-source and flexible platform for low-end VLC research, we introduced OpenVLC at the VLCS'14
workshop~\cite{Wang2014vlcs}, that allowed for quick and flexible testing of new VLC protocols and applications. More recently, 
we introduced OpenVLC1.2~\cite{openvlc12} with the attempt of increasing the data rate. However, the board was still working only at relative 
short range and was largely affected by light interference.

In this paper, we introduce the latest version of our platform, OpenVLC1.3, that increases the data
rate and the communication range without adding any hardware cost to the platform. Our contributions are as follows:
\change[Ander]{Domenico, trackchange doesn't like working with itemize elements, so I'm going to change the order of this without using it}{}
\begin{itemize}
\item We make a design that occupies a smaller physical space, improve the hardware of OpenVLC1.3 and add high-pass and low-pass filters to minimize the effect of noise in the system, including the Direct Current (DC) from other illumination sources and high-frequency components from the circuitry such as the overshooting generated by the amplification stages.
 \item We improve the software stability and make a new design in software to modulate the LED light. This allows us to perform sampling rate over 2\,MHz and achieve a UDP throughput of \change[Ander]{above}{about} 400 kb/s. This throughput could fulfill the needs of a range of applications with only off-the-shelf low-end hardware.
\item \change[Ander]{Since the PHY layer runs in a 
	memory-constrained microcontroller, we also design a new technique for 
	computation- and memory-limited fast frame detection.} {We design a new technique for computation- and memory-limited fast frame detection to solve the problem of constrained memory in the microcontroller.}
\item We design and implement a new reception mechanism to avoid the synchronization problems present in previous 
versions of OpenVLC.
\end{itemize}

\begin{figure}
	\centering
	\includegraphics[width=.5\columnwidth]{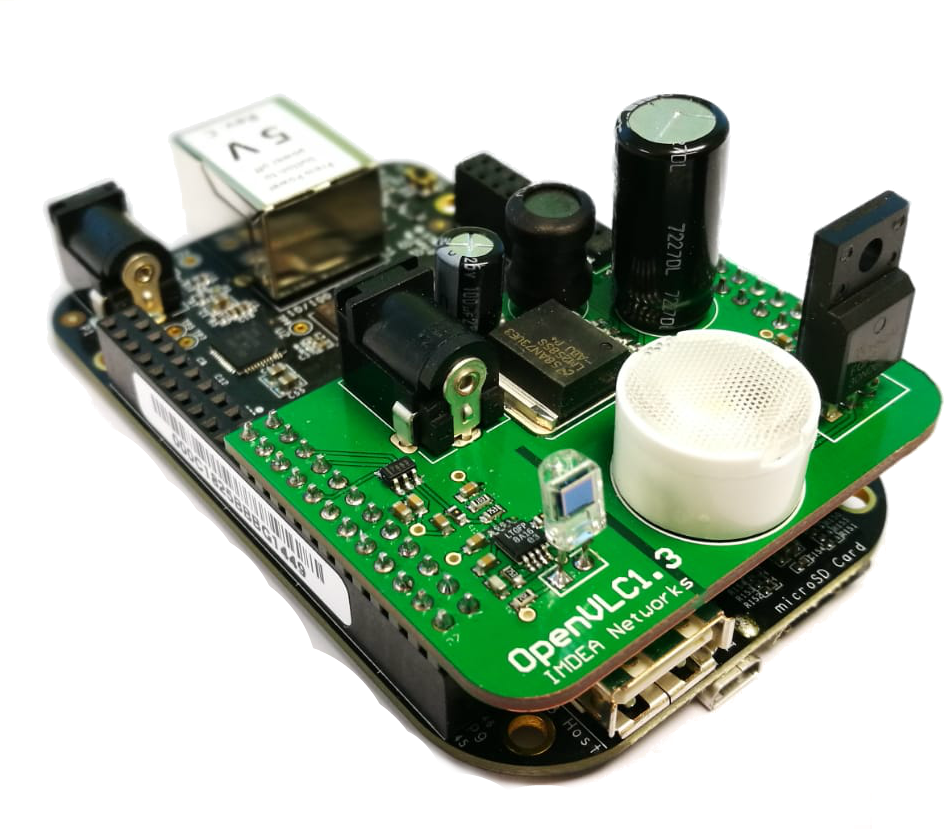}
	\captionof{figure}{OpenVLC1.3 cape on top of an embedded board.}\label{fig_cape}
\end{figure}

The rest of this paper is organized as follows. Section~\ref{sec_arch} introduces the background on OpenVLC and a high level view 
of the system architecture. Details on the design of hardware, firmware and driver for the transmitter (TX) and 
receiver (RX) are presented in Section~\ref{sec_tx} and Section~\ref{sec_rx}, respectively. The evaluation results are reported in 
Section~\ref{sec_eval} and the limits of the OpenVLC1.3 in Section~\ref{sec_limits}. Finally, discussions and 
conclusion are drawn in Section~\ref{sec_conclusion}.

\section{New System Architecture}  \label{sec_arch}

\begin{table*}[t]
	\centering
	\caption{Comparison between OpenVLC versions.}
	\label{table_versions}
	\begin{tabular}{|c|c|c|c|c|c|}
		\hline
		 & TX HW                                                                                         & TX SW                                                                                      & RX HW                                                                                          & RX SW                                                                                          & Data rate \\ \hline
		OpenVLC1.0      & High Power LED                                                                                & Kernel software                                                                            & Basic components                                                                               & Running in Kernel                                                                              & 18kb/s    \\ \hline
		OpenVLC1.2      & \begin{tabular}[c]{@{}c@{}}Support for higher power \\ LED and faster modulation\end{tabular} & \begin{tabular}[c]{@{}c@{}}Firmware with user space\\ connection\end{tabular}              & \begin{tabular}[c]{@{}c@{}}Faster PD and\\ external amplifier\end{tabular}                         & \begin{tabular}[c]{@{}c@{}}New frame detection and\\ faster reception in firmware\end{tabular} & 100kb/s   \\ \hline
		OpenVLC1.3      & \begin{tabular}[c]{@{}c@{}}More powerful LED \\ and external power\end{tabular}               & \begin{tabular}[c]{@{}c@{}}Faster firmware and \\ direct connection to Kernel\end{tabular} & \begin{tabular}[c]{@{}c@{}}Filters to remove interferences\\ and reduced cape size\end{tabular} & \begin{tabular}[c]{@{}c@{}}New frame and \\ symbol detection \end{tabular}              & 400kb/s   \\ \hline
	\end{tabular}
\end{table*}

\begin{figure}[t]
	\centering
	\includegraphics[width=0.8\columnwidth]{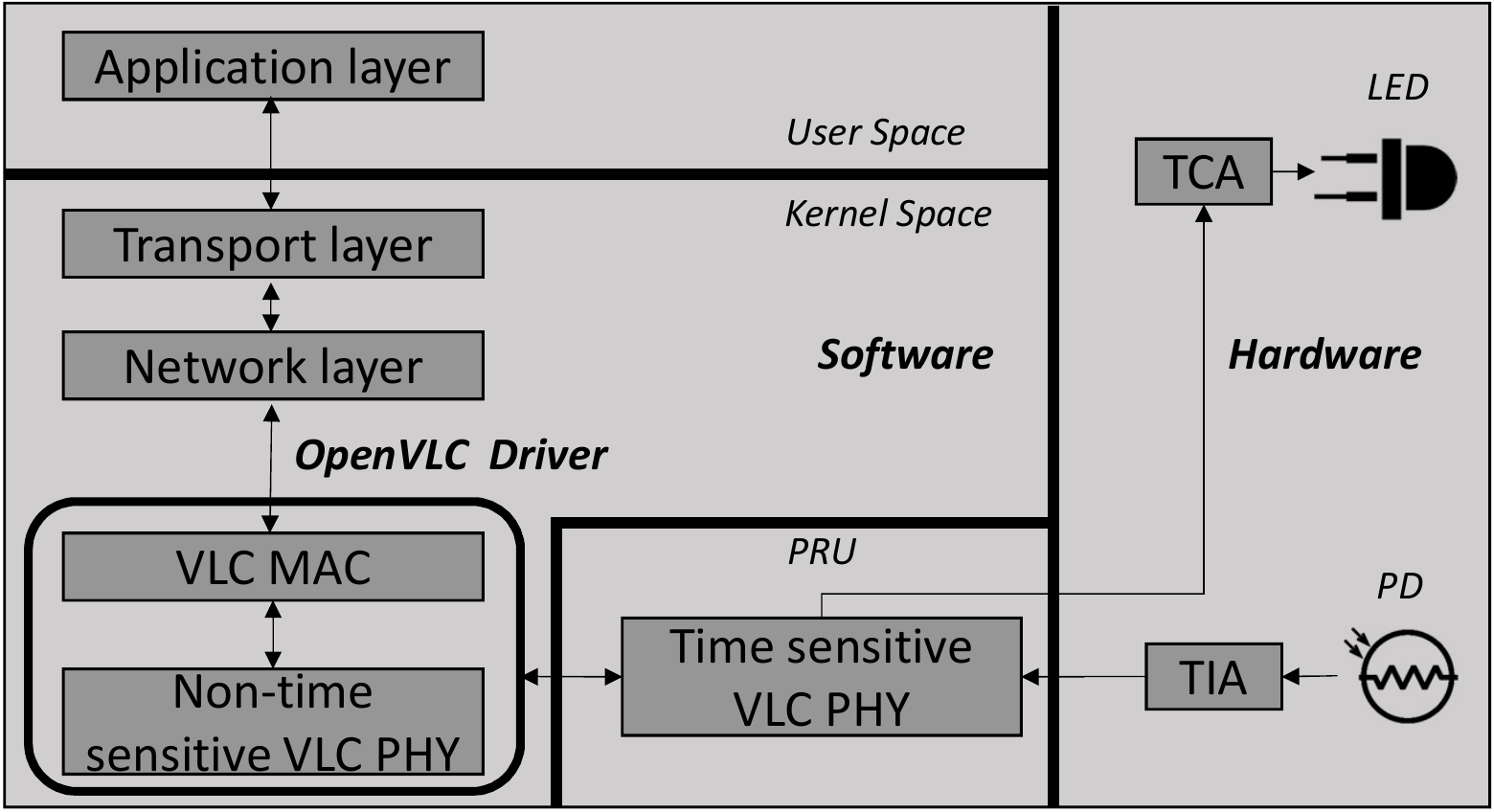}
	\caption{The diagram of OpenVLC1.3.}
	\label{fig_diagram}
\end{figure}

The \change[Ander]{last}{new} version of OpenVLC  consist of four parts: the BeagleBone Black (BBB) embedded board~\cite{BBB}, the OpenVLC 
cape, the OpenVLC firmware and the OpenVLC driver. The OpenVLC cape is the front-end transceiver that is attached directly to the 
BBB. \change[Ander]{The OpenVLC firmware and driver implement the real-time processing in the BBB's Processing Real-time Units (PRUs) and the VLC MAC and PHY layers in the Linux kernel, and}{The OpenVLC firmware uses real-time processing in the BBB's Processing Real-time Units (PRUs), that work as microprocessors. The OpenVLC driver is a module in the Linux kernel. Both the firmware and the kernel module implement the VLC MAC and PHY layers and} implement primitives such as sampling, 
symbol detection, coding/decoding and Internet protocol interoperability. Also, OpenVLC1.3 retains the best characteristics of previous versions, being flexible and open-source and \change[Ander]{being equipped}{communicating} with a low-cost front-end.

\change[Ander]{Based on its predecessor}{With respect to its predecessor}, the architecture of OpenVLC1.3 has been re-designed in order to increase the network performance.
The new hardware (HW), called OpenVLC1.3 cape, is shown in Fig.~\ref{fig_cape}. The OpenVLC1.3 cape has been modified, reducing its surface by more than 50\%. This 
also allows to use the remaining pins to connect 
sensors, for instance for Internet of Things applications. 

The system architecture of OpenVLC1.3 is shown in Fig.~\ref{fig_diagram}. The hardware runs on external power to allow higher power consumptions and harnesses the new LED and 
Photodiode (PD), together with ancillary circuits, to transmit and receive visible light signals, respectively. The software is 
responsible for modulating \change[Ander]{the LED light to transmit and sampling the incoming signals}{the LED light in order to transmit and sample the incoming signals} to receive, both implemented 
in the OpenVLC1.3 firmware. The software also implements the MAC layer and part of PHY layer in the OpenVLC1.3 driver.

There are three main differences in the design comparing OpenVLC1.3 to its predecessors:

\begin{itemize}
	\item  \change[Ander]{An improved}{A new design of the} OpenVLC cape (hardware).
	\item A new system architecture (both in software and hardware). 
	\item A firmware implemented in the PRUs for data transmission as well as frame and symbol detection (software).
\end{itemize}

To boost the date rate in OpenVLC1.3, \change[Ander]{we exploit the BBB's.}{we exploit the PRUs of the BBB.} 
Time-sensitive operations are implemented in the \change[Ander]{}{2} PRUs \change[Ander]{(OpenVLC firmware)}{} that control the General Purpose Input-Output (GPIO) to 
modulate LED light and perform sampling of incoming signals. This separation was also proposed in OpenVLC1.2, but resulted in 
overall lower performance and required some module in user space. Communication between the driver and the firmware is \change[Ander]{}{now} performed using a shared memory. A new technique for 
computation- and memory-limited frame detection also resides in the firmware (the details are presented 
in Sec.~\ref{sec_rx}). The OpenVLC driver implements the MAC protocol and non-time sensitive PHY operations. This maintains 
the advantages of software-based flexibility and programmability while increasing its performance.

OpenVLC1.3 is already available to the research community\footnote{\url{www.openvlc.org}}.
A summary of the improvements of each version can be found in Tab.~\ref{table_versions}.

\subsection{Data exchange}

The data stream is received in the driver from upper layers. The VLC frame is prepared and then the symbol stream is sent to 
the shared memory from where it is read by the 
firmware in the PRU, as seen in Fig.~\ref{fig_diagram}. The PRU then controls the GPIOs to modulate the LED light for data transmission. 

At the receiver, light signals are detected by the PD and sampled by the firmware in the PRUs. 
Once a valid preamble and \change[Ander]{SFD}{Start-Frame-Delimiter (SFD)} are detected, received data is sent to the shared memory, \change[Ander]{which then is read}{and then received} and processed by the OpenVLC 
driver. Finally, the received data is sent to the network layer, where it is handled using the TPC/IP Linux kernel. 

\begin{table*}[t]
	\centering
	\caption{Frame format and sizes (in bytes).} 
	\label{table_frame}
	\resizebox{1.45\columnwidth}{!}{
		\begin{tabular}{|c|c|c|c|c|c|c|}
			\hline
			 Preamble & SFD & Frame Length & Dst. Address & Src. address  & Payload & Reed-Solomon \\ \hline
			 3 & 1 & 2 & 2 & 2  & 0-MAX & 16\\ \hline
	\end{tabular} }
\end{table*}

			
\subsection{Firmware} \label{sec_firmware}

The firmware of OpenVLC runs in the PRUs of the BBB, which operates at 200\,MHz, meaning that each instruction takes 5\,ns. Each PRU has its own memory and a shared one between the two. The size of each memory of the PRUs is 8KB and the shared memory is 12KB.
The reason behind adopting the PRUs in OpenVLC is to increase the data rate and handle a higher sampling frequency of the 
Analog-to-Digital Converter (ADC). Nevertheless, 
this effort also requires a tight timing precision in both the modulation and sampling processes.  For this reason, 
\emph{assembly} is used 
to program the PRU. In this way, the code of the PRU has been designed to know the exact number of 
instructions executed and, subsequently, the time required to execute them. In addition, the memory space in the 
PRU is very limited and this requires careful optimization of all instructions. Finally, there is no enough memory 
to implement queues, and as such, the communication between the PRU and upper layers must be handled carefully.

\subsection{Kernel Driver}\label{sec_driver}

The main objective of OpenVLC is to have a flexible, low-cost and reconfigurable system\change[Ander]{}{ for communication using visible light}. In order to do so, 
OpenVLC1.3 has been designed to be as
versatile and easy to use as possible. For this reason, we have taken two design decisions:
\begin{itemize}
	\item OpenVLC is mostly code-based and the use of VLC hardware is as small as possible. This makes easy to modify the 
	behavior of the platform just by modifying the\change[Ander]{}{ software} code, such as introducing new MAC protocols.
	\item OpenVLC should be easy to use and adaptable to most use case scenarios. Taking this into account, OpenVLC's 
interface has been designed as a Linux kernel module.
\end{itemize} 

The OpenVLC \change[Ander]{}{kernel module} allows us to create a network interface. This means that any user will see the OpenVLC module as 
if it is just another network device such as Wi-Fi or Ethernet and any application that we would like to run would be connected 
through the VLC network interface. As the kernel runs in the processor of the BBB, its processing power is much higher than the 
one of the PRU\change[Ander]{}{ microcontroller}. For this reason, the most computationally demanding tasks are left in the kernel.




\section{Transmitter}  \label{sec_tx}
In this section we \change[Ander]{show}{present} all the different parts that allow OpenVLC to perform a VLC transmission.


\subsection{Kernel module for transmission}

When a user space application transmits data, first the packet is received from the IP layer of the kernel.
After unwrapping the frame, the driver 
prepares the header for the VLC MAC layer. The frame structure 
is presented in Table~\ref{table_frame}. Each frame starts with a frame 
header that contains the following fields: preamble, SFD, frame length, destination address and source 
address. Each field in the MAC header (starting from the frame length) can be freely modified, for instance to 
adapt it to the IEEE 802.15.7-2011 standard for VLC~\cite{8021157}\change[Ander]{}{ and upcoming 802.11bb standard for integration with Wi-Fi}\cite{802_11bb}. 

The preamble consists of 24 alternating HIGH and LOW symbols. After that, the SFD is appended to avoid false positives. The next 
field denotes the length of frame in bytes, followed by the destination and source addresses.

We use Reed-Solomon code to correct errors in the data field during the transmission. The bits for Reed-Solomon are appended to 
the frame. We use Reed-Solomon (216,200) error correction code in our default configuration. Subsequently, we use Manchester line 
encoding, which encodes one bit into two symbols with On-Off-Keying (OOK) modulation
(a symbol is either a HIGH or a LOW) and it 
ensures that the average 
signal power remains constant. In particular, Manchester line coding converts a 1 bit into a LOW-HIGH symbol pair and a 0 into a 
HIGH-LOW.  This is done to avoid flickering. 
Both Reed-Solomon code and Manchester 
encoding are also used in the 802.15.7-2011 standard~\cite{8021157}. Finally, 
the driver places the VLC frame in a shared memory, so that the OpenVLC firmware\change[Ander]{ can access it from the PRUs}{ in the PRU can access it}.

\subsection{Shared memory}

\change[Ander]{How the memory is shared can be seen in Fig.~\ref{fig_memory_rxa}.}{In Fig.~\ref{fig_memory_rxa} we show how the shared memory is used in OpenVLC. }The kernel driver transmit data to the PRU using a shared memory. The first 32-bits {(referred as the first `register' in the rest of 
	this paper) of the shared memory are the only space where the PRU can write data. \change[Ander]{It is this register where the kernel 
	checks if the PRU has finished with the transmission}{This register is used to exchange status flags between the kernel and the PRU}.
	
	The communication works as follows: the PRU constantly reads the value of the first register. If it is zero, it reads it again in 
	a loop, waiting for its value to change. If the kernel receives data from upper layers, it will modulate it and put it the shared 
	memory. When it finishes, it writes  in the first memory address the number of registers that the 
	PRU should read. The kernel will not be able to write into the shared memory again until the PRU finishes\change[Ander]{}{ the transmission}.
	The PRU will then transmit the data and once it finishes, it changes the flag in the first memory register to zero so that the kernel 
	knows that the memory is available again.


	%
	%

	\begin{figure}[t]
		\centering
		\subfigure[Memory sharing between the OpenVLC driver and the PRU$_0$ at the transmitter.] 
{\includegraphics[width=.25\columnwidth]{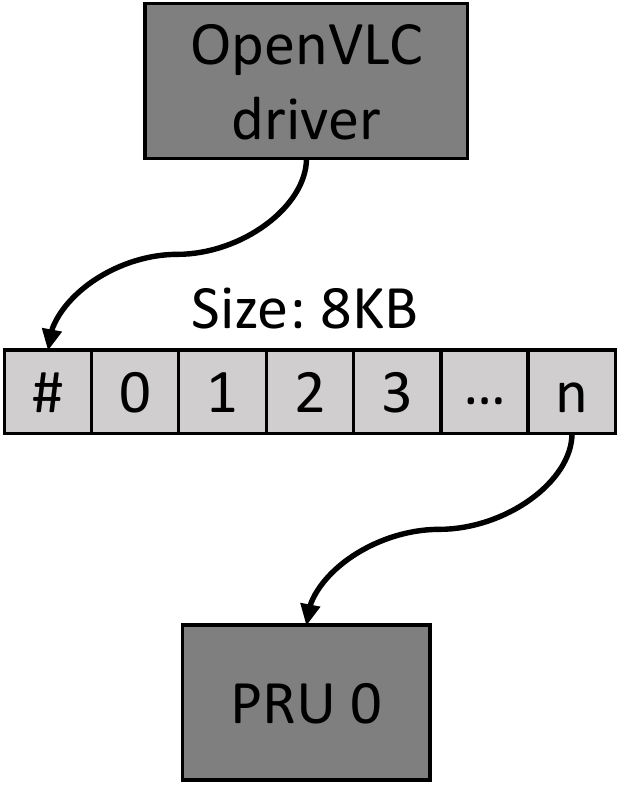}
		\label{fig_memory_rxa}}
		\hspace{5mm}
		\subfigure[Memory sharing between the PRUs and between PRU$_1$ and the OpenVLC driver at the receiver.] 
{\includegraphics[width=0.60\columnwidth]{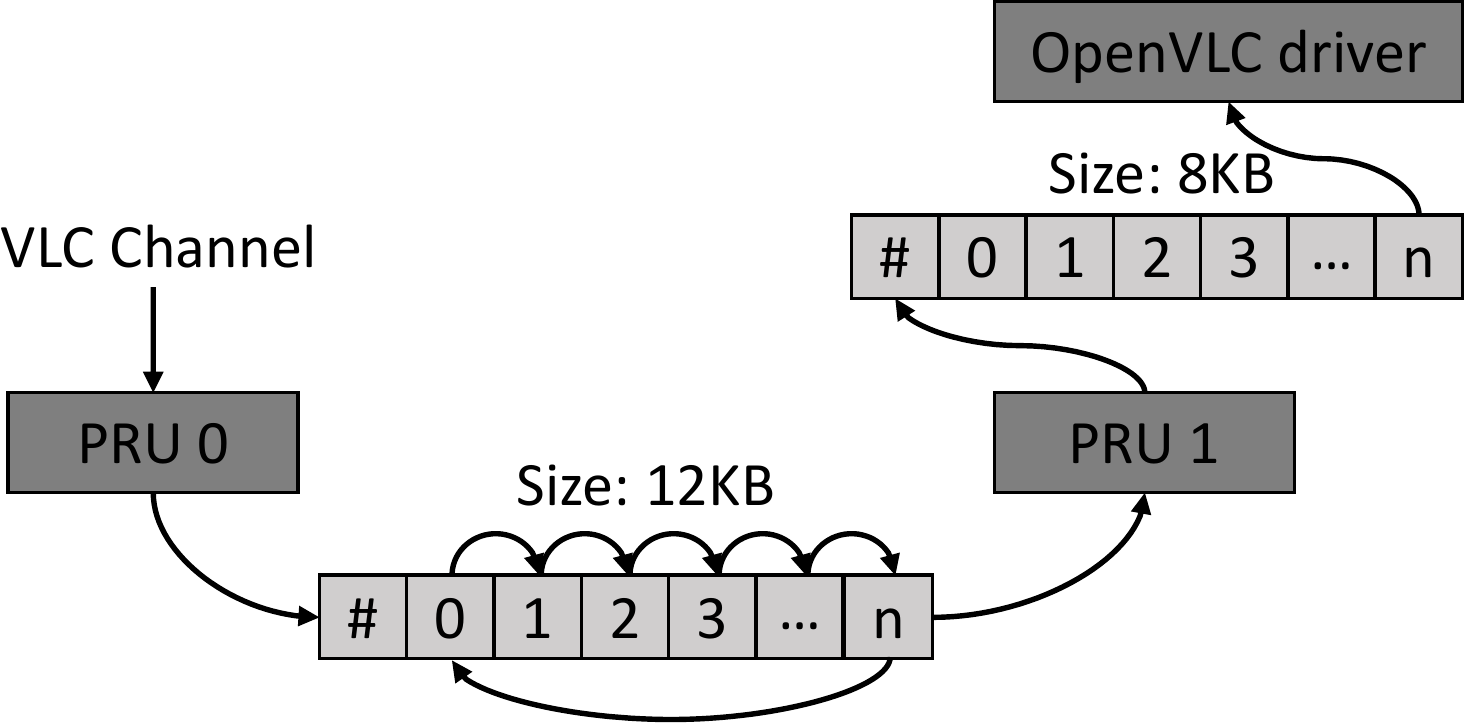}
		\label{fig_memory_rxb}}
		\caption{Memory interface of OpenVLC1.3 (`\#' stores the physical address of the latest updated data).	} 
		\label{fig_memory_rx}
	\end{figure}

\subsection{Firmware for signal transmission}
\label{sec_firm_tx}

For the transmission, the PRU is used for the sole purpose of emitting the visible light signal according to the
pattern of HIGH and LOW symbols stored in the shared memory. Two PRUs are available in the BBB, 
and only one PRU is used for transmitting the signal. We implement a counter   to track the time between symbols. When it 
reaches zero, a new symbol is transmitted. 
In order to transmit each symbol, a HIGH or LOW signal is sent through one of the pins of the BBB.
In the current implementation, HIGH corresponds to emitting the visible light signal and LOW to not emitting
any signal.

\subsection{Hardware improvement}

The main purpose of the TX circuit is to take the signal given by the BBB and amplify it to \change[Ander]{turn on/off the LED}{turn the LED on/off}. As mentioned 
above, the HW is controlled using the firmware implemented in the BBB's PRUs. The HW used for the OpenVLC transmitter can be 
seen in 
Tab.~\ref{table_components}.
We have improved the TX circuit compared with previous versions mainly to support a higher transmission rate and a 
larger communication range. 

\begin{itemize}
	\item \emph{Increase the transmission rate:} we use a PRU at the TX to modulate the LED light at higher speeds. 
	Moreover, we use a MOSFET gate driver transistor to control the current flowing to the LED and support a faster switch.
	\item \emph{To expand the communication range:} a LED that maximizes the power supported by the transmission circuit, working at 2.8\,W with a luminous flux of 400\,lm. A lens has been attached on top of the LED to \change[Ander]{concentrate more}{better concentrate} the optical power and thus, reach further distance. A heat-sink is attached to better dissipate the heat generated by the high-power LED.
\end{itemize}

\section{Receiver}  \label{sec_rx}
In this section we explain the design and behavior of the~RX.

\subsection{Hardware for reception}
In the previous versions of OpenVLC, the bottleneck for the throughput was the RX's sampling rate. In OpenVLC1.3, this is 
solved partly by 
introducing a new faster photodiode (PD). This PD does not have its own amplifying circuit. Thus, we add an external amplifier to 
the RX. The PD's position in the cape is also adjusted for a better detection of visible light. The most important 
components are shown in Tab.~\ref{table_components}.

\begin{table}[]
	\centering
	\caption{Main components of OpenVLC1.3.} 
	\label{table_components}
	\begin{tabular}{|c|c|}
		\hline
		
		\textbf{Component} & \textbf{Name} \\	\hline \hline
		ADC       & ADS7883  \\ \hline
		OP-AMP    & LTC6269  \\ \hline
		MOSFET    & FQPF30N06L \\ \hline
		LED       & XHP35A-01-0000-0D0HC40E7CT  \\ \hline
		Lens      & TINA FA10645  \\ \hline
		PD        & SFH206K  \\ \hline
		DC-DC     & LM2585SX-ADJ \\ \hline
	\end{tabular}
\end{table}

The bottleneck of the system for the transmission distance on the reception circuit was the high sensitivity to noise 
on the receiver circuit. For this 
reason, a reception chain has been added between the PD and the ADC. In previous versions there was only an amplification 
stage between the PD and the ADC. In this version, as seen in Fig.~\ref{fig_rxchain}, the first amplification stage is a 
low-noise Trans-Impedance-Amplifier (TIA) that converts the current of the PD into voltage. 

Subsequently, a high-pass filter is used in order to remove the low frequency components (specially the DC component 
from other illumination sources). The cut-off 
frequency of this filter is 10\,KHz. This filter allows us to remove the DC light component and other sources of interference. 
Although non-visible for the human eye, a light flickering at this frequency would distort the VLC signals. After this, a DC component of 2.5\,V is added to the signal so that the signal is centered at half the span of the ADC.
Then, the second amplification stage prepares the signal for the dynamic range of the ADC. Finally, before the ADC, a low-pass 
filter with 
cut-off frequency of 1.1\,MHz removes the higher frequency noise components mainly due to overshooting of the amplifiers.

\subsection{Firmware for signal reception}

The configuration \change[Ander]{on}{of} the RX is more complex than \change[Ander]{on}{of} the TX. It requires two PRUs. One for handling the HW in a very precise manner and another for processing the received signal. 

One of them, PRU$_0$, performs signal sampling from the ADC and obtains the Received Signal Strength (RSS) and sends it to the 
other PRU, PRU$_1$, that handles signal detection and \change[Ander]{bit extraction}{the process of converting the raw signal into bits}.
PRU$_0$ reads the RSSs from the ADC at a frequency higher than twice the symbol rate. \change[Ander]{In this way, the system assures that even if the system gets de-synchronized, it will get at least 1 sample per symbol.}{} 

Then, the raw value from the channel is shared with PRU$_1$. PRU$_1$ interprets the RSSs into symbols for frame 
detection. PRU$_1$ continuously checks if a new RSS has been read by PRU$_0$. If yes, PRU$_1$ processes it immediately.

\subsubsection{The bit slip problem}
One of the most sensitive stages in a communication system  is the correct reception of transmitted symbols. One of the main 
problems with low-cost systems is that TX and RX get\change[Ander]{}{ easily} desynchronized over time. Their clocks are not exactly the same and the frequencies at which they run are slightly different. This could make the system sample a symbol twice or miss a symbol. 
This problem is known as  ``bit slip'' \cite{bitslip}.
In\change[Ander]{}{ the older} OpenVLC1.2, the TX and RX frequency \change[Ander]{are}{were} adjusted to the \textit{instruction level}. This meant that there are exactly the 
same number of instructions between two symbols transmission and between two symbols reading. However, the clocks in the TX and 
RX always run at slightly differently frequency, and thus, part of the 
synchronization problems was still present. 

In order to solve this problem, we need to make sure that:

\begin{itemize}
	\item All the symbols \change[Ander]{need be sampled}{are sampled} at least twice.
	\item The system should detect if a symbol has been sampled more than twice.
\end{itemize}

\change[Ander]{T}{OpenVLC1.3 over-samples the signal t}o assure that all the symbols are sampled at least twice\change[Ander]{, OpenVLC1.3 over-samples the signal}{}. The higher the oversampling rate, the more information the system is going to have to detect the symbol correctly. 
Nevertheless, a high oversampling rate requires a fast processing. In our case, in order to fulfill the requirements \change[Ander]{above mentioned in OpenVLC1.3}{mentioned above}, the sampling frequency $f_{sampling}$ should be:

\begin{figure}[t]
	\centering
	\includegraphics[width=0.9\columnwidth]{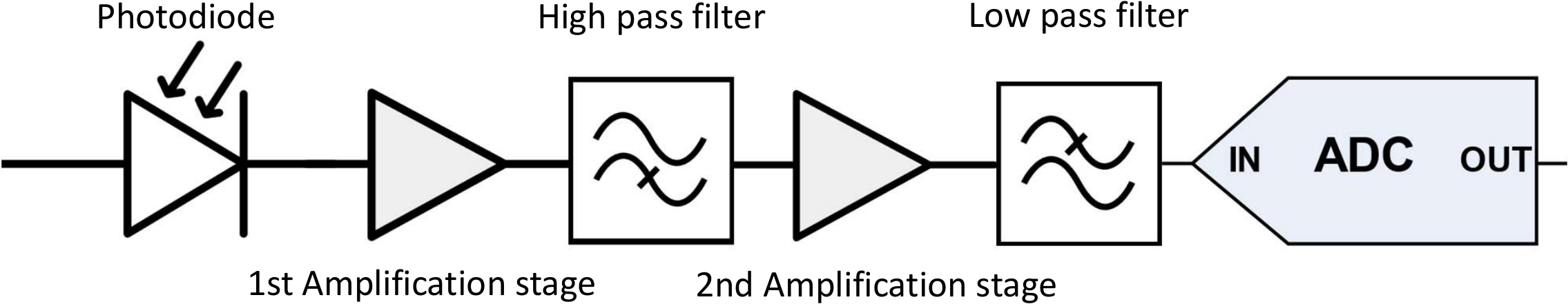}
	\caption{Reception chain.}
	\label{fig_rxchain}
\end{figure}

\begin{equation}
2f_{symbol}<f_{sampling}<3f_{symbol}
\end{equation}

With this configuration, OpenVLC1.3 makes sure that we always receive at least 2 sample per symbol and a maximum of 3. 
It is not possible to receive 4 samples per symbol, which is \change[Ander]{really important}{necessary} to assure the second condition.
In the implementation, we modulate at 1\,MHz and sample at the receiver at a rate of 2.1\,MHz.


The symbol detection in OpenVLC1.2 was just a thresholding algorithm with one sample per symbol with the consequent bit sleep 
problem. In OpenVLC1.3, we avoid this by using a pseudo-edge detection algorithm.
Manchester modulation converts a 1 bit into a LOW-HIGH symbol pair and a 0 into a HIGH-LOW. This means that the 
maximum number of symbols with the same value is 2. There cannot be more than 2 consecutive HIGHs or LOWs.
With two samples per symbol, the system makes sure to read at least each symbol twice, at 
it can be seen in Fig.~\ref{fig_desynch}.

OpenVLC1.3 can count the number of samples with the same value. If the number of samples is 2 or 3, only 1 symbol has been 
received. If the number of samples is 4 or 5, 2 symbols have been received. This method can be thought as a very simple and 
rudimentary edge detection system, as it looks for changes in the signal to see when a new symbol (or pair of symbols with the 
same value, depending on how long ago was the last change in the signal) has been received.

\begin{figure}[t]
	\centering
	\subfigure[1 sample per symbol (symbol lost).] {\includegraphics[width=0.4\columnwidth]{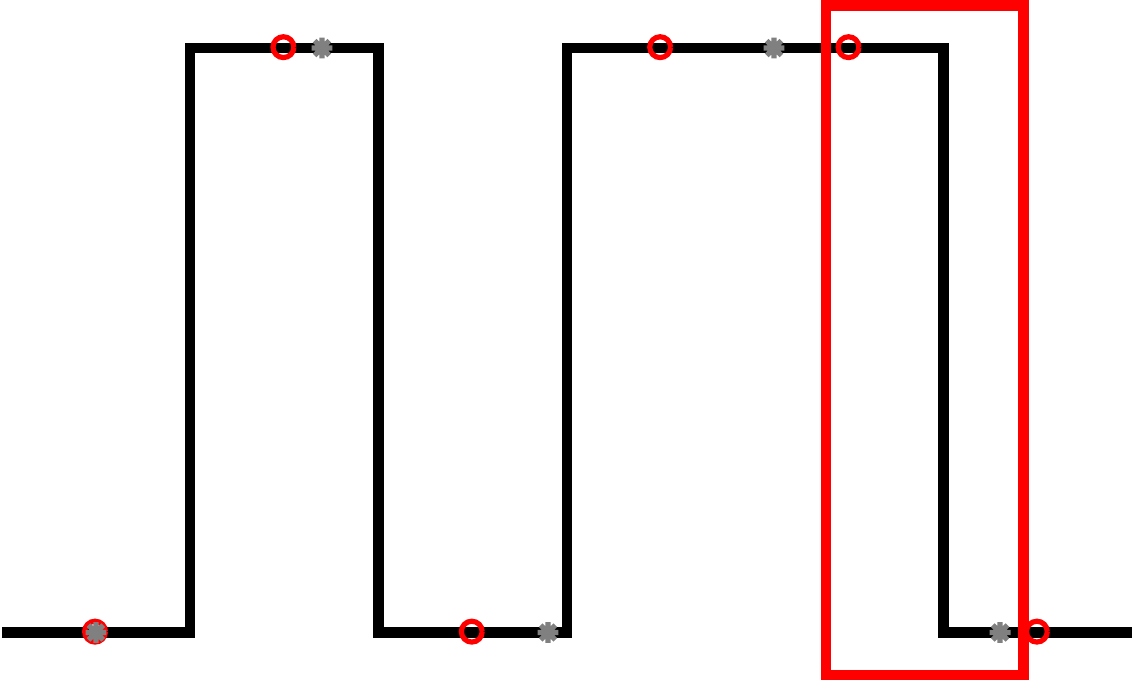}}
	\hspace{5mm}
	\subfigure[2 sample per symbol (no symbol lost).] {\includegraphics[width=0.4\columnwidth]{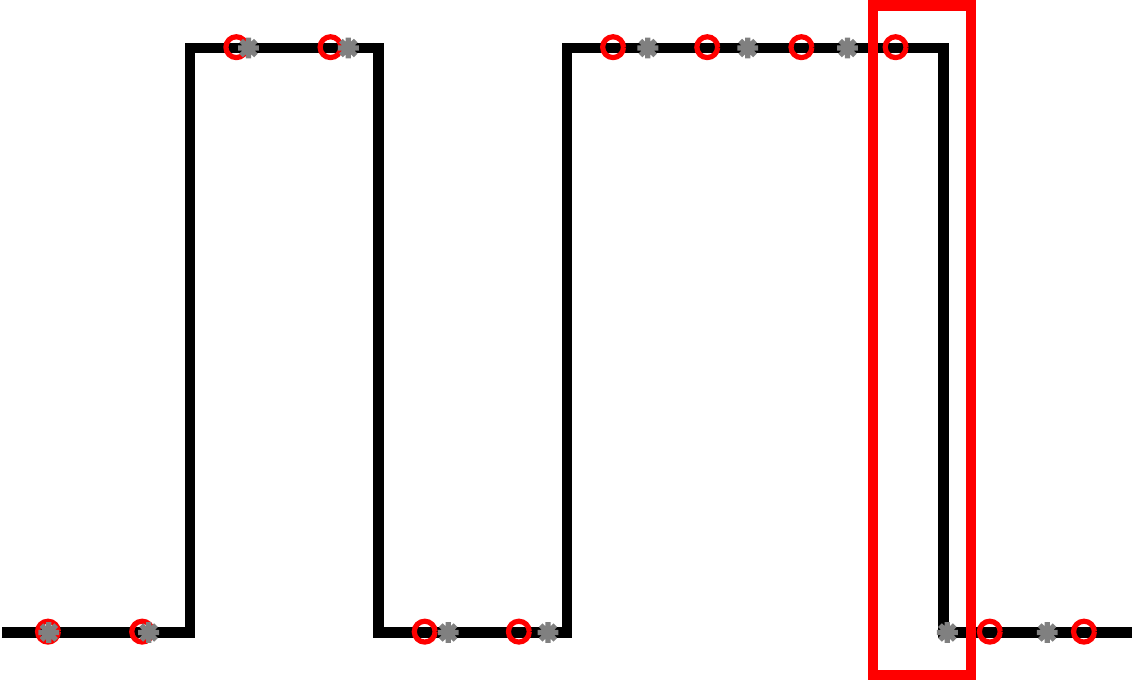}}
	\caption{Bit sleep problem. The red circles represent the ideal sampling time. The grey stars represent the  real sampling time given by the drift.} \label{fig_desynch}
\end{figure}

\subsubsection{Frame detection}
The system assumes that a new frame is being detected when the last samples received correspond to the preamble + SFD. The frame 
detection technique is the one used in OpenVLC1.2 \cite{openvlc12}
as it showed to be both \change[Ander]{light}{low-complexity} in processing (required by our application) and very effective.
This technique works as follows: First, the RSS values read are compared with their previous values. 
Because at the beginning of a frame in the preamble (0xAAAAAA) 
every HIGH is between two LOWs and every LOWs between two HIGHs, every symbol is different from the previous one.  If 
the value of the last 24 bits received is not the same as the preamble, the system continues to collect samples. If it does, continues receiving data.

Once the preamble is detected it continues receiving the rest of the frame. Once received and demodulated, the data is sent to the 
OpenVLC driver for further processing.

\subsubsection{Communication between PRUs}

The signal reception starts with the PRU$_0$ reading values from the ADC and putting them in a memory shared by both PRUs. 
This memory is used as a circular memory. When it reaches the end, it continues filling the beginning of the shared memory. In 
the first register, the address of the latest memory where data has been written is stored as illustrated in 
Fig.~\ref{fig_memory_rxb}. In this way, PRU$_1$ is able to keep track of the RSS obtained by PRU$_0$ in real time. Then, 
the 
PRU$_1$ processes the samples taking two symbols at a time, to determine if the encoded Manchester bit is a 1 or a 0. If they contain \change[Ander]{valuable}{valid} data, it is decoded and shared with the kernel using the same process as for 
the transmitter.


%
%
%
%
\subsection{Kernel for reception}
\change[Ander]{The packet will be directly received in the kernel after being converted from symbols to bits 
	by the PRU}{The frame is received by the kernel after being converted from symbols to bits in the PRU}. The Reed-Solomon code is checked with three possible outputs:

\begin{itemize}
	\item The Reed-Solomon reports no errors, so the packet is forwarded.
	\item The Reed-Solomon shows some errors that is able to correct, so corrects them and forwards the packet.
	\item The Reed-Solomon code shows that there are too many errors, so discards the packet.
\end{itemize}

If the packet is forwarded, the kernel encapsulates the packet so that upper layers can manage it.

Both in the transmission and in the reception of packets, OpenVLC considers that, although more powerful, 
the kernel cannot run \change[Ander]{on}{in} real time. For this reason, two driver queues are implemented, one for transmission and one for 
reception. Every incoming packet that arrives to the kernel is queued and transmitted to the PRU or to upper layers as soon as 
the resources are available. \change[Ander]{This way, the chance to lose a frame because the CPU is occupied is minimized.}{In this way, we minimize the likelihood of losing a frame because the CPU is occupied.}
\section{Evaluation} \label{sec_eval}

In this section we evaluate the performance of OpenVLC1.3.

\begin{figure*}[htp]
	
	\centering
	\includegraphics[width=.26\textwidth]{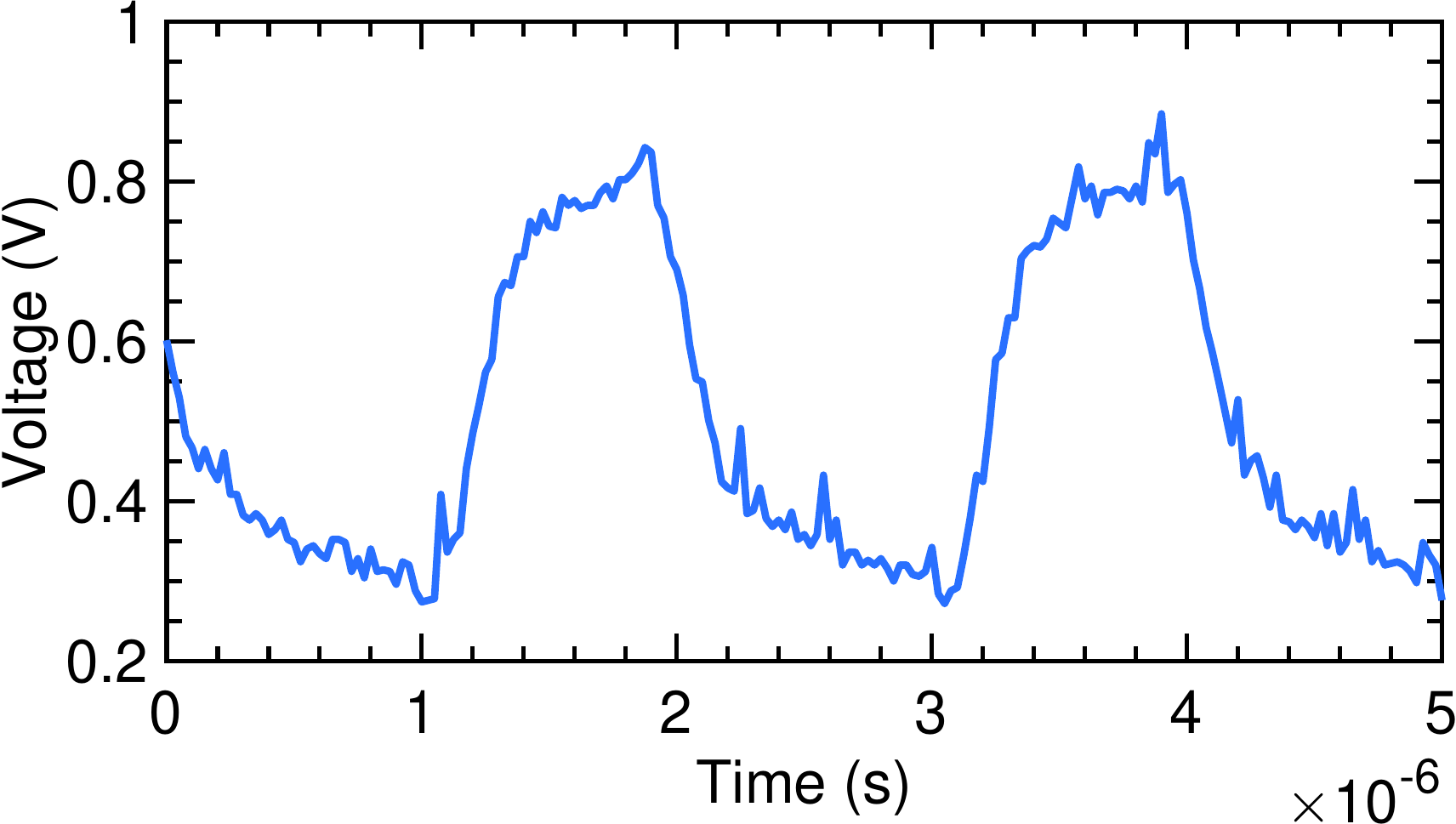}\hspace{10mm}
	\includegraphics[width=.26\textwidth]{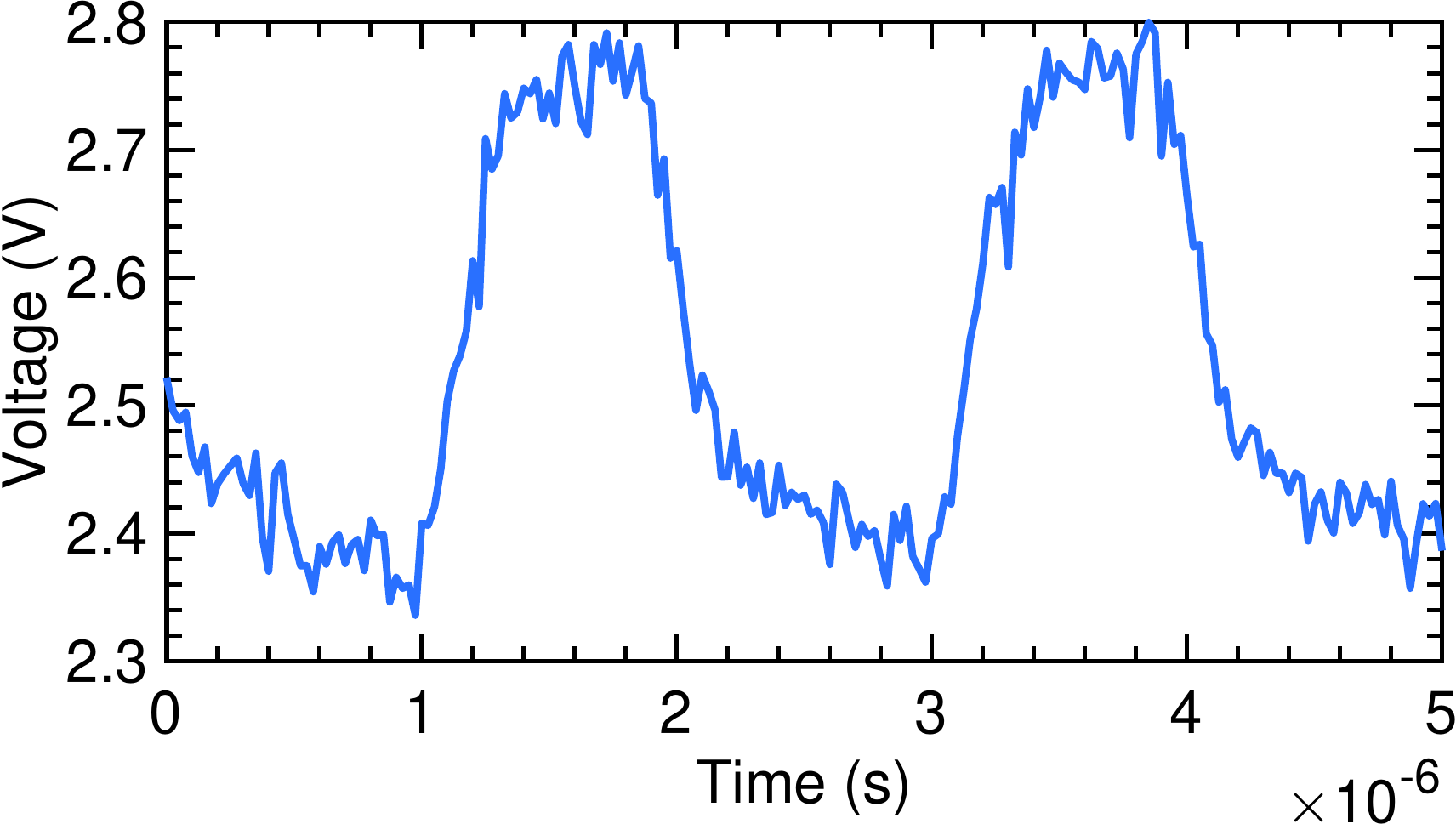}\hspace{10mm}
	\includegraphics[width=.253\textwidth]{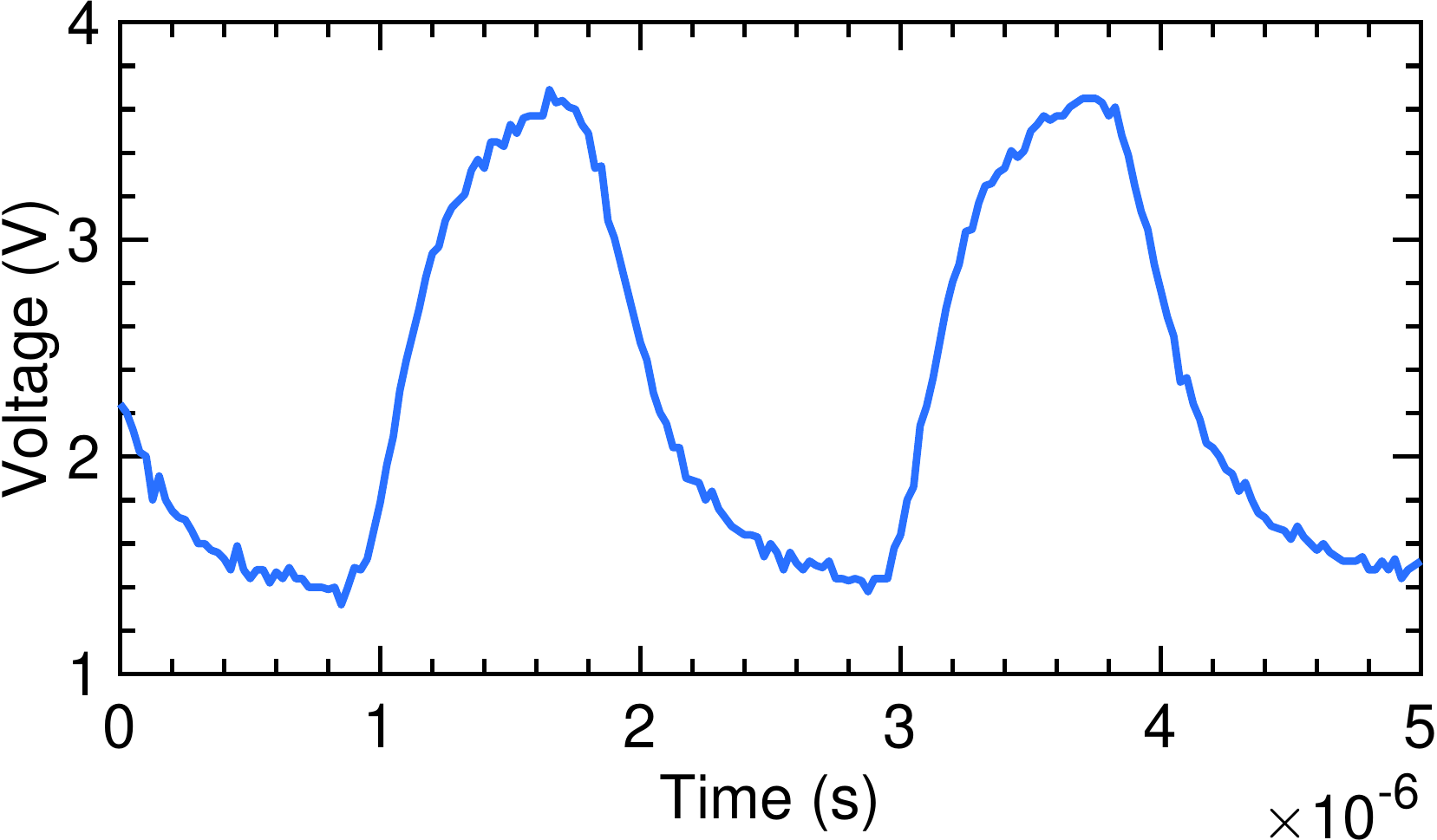}
	
	\caption{Received signal after the first amplifier (left), after the high-pass filter (center) and at the entrance of the ADC (right).}
	\label{fig_chain_meas}
	
\end{figure*}

\subsection{Reception chain}

In order to understand the behavior of the reception chain we have measured the raw signal with an oscilloscope. As it can be seen in Fig.~\ref{fig_chain_meas},\change[Ander]{ after the first amplifier}{} the signal is noisy and small\change[Ander]{}{ after the first amplifier}. Then, the signal is filtered and centered around the center of the ADC's span. Finally, the signal is amplified and cleaned, to improve the reception.

\subsection{Throughput vs. payload}

\subsubsection{Setup} We use two OpenVLC1.3 nodes, one as TX and the other as a RX.\change[Ander]{ Although technically possible to use OpenVLC 
	as a transceiver, in the current version, the throughput is maximized when the resources are focused on only 
	transmitting or receiving.}{} Since OpenVLC1.3 provides a new network interface that can be 
easily accessed by upper layer applications, we use the tool \emph{iperf} to evaluate the UDP performance of OpenVLC1.3. 

\subsubsection{Results} The first test performed is to see how the system works depending on its payload. In previous 
versions of OpenVLC, the payload had a huge impact in the system. If the payload was too short, the overhead due to the physical 
layer headers was too big, decreasing the throughput. Nevertheless, if the payload was too big, the reception was desynchronized 
and the frame lost. \change[Ander]{Anyway, as in OpenVLC1.3 the symbol detection technique is modified}{As we modify the symbol detection technique}, now no frames are lost due to 
the size of the payload. For this reason, the bigger the payload, the better. All the following tests are done with payloads of 800 bytes.

\textit{Throughput vs. distance.} This test contains the two most important parameters regarding VLC. The first one is the 
distance at which the VLC communication takes place. The second one is the maximum throughput achievable by the system. 
OpenVLC has been tested over distance under 3 different scenarios. In the first one, the system has been deployed in a realistic 
scenario with no artificial lights on (here OpenVLC is seen \change{to be}{as} the primary illumination source), but with the windows open 
during the day (``W. open''). In the second one, OpenVLC is tested without any external light interference and the window shutter 
closed (``W. 
closed''). In the last one, we open the windows, and we add an artificial fluorescent light source with frequency 
components that enter in the frequency band of the OpenVLC receiver (``Interference'').

As it can be seen in Fig.~\ref{fig_th_payload}, the maximum throughput that OpenVLC achieves is 400\,kb/s. 
Until 3.5 meters, the 
difference between being in a completely dark environment (``W. 
closed'') and with external light (``W. open'') is negligible, which did not happen in previous 
versions. This is due to the filters added in the reception chain. Then, at 3.5 meters the intensity of the external light level 
becomes similar to transmitted light, making it more difficult for the ADC to differentiate between HIGH and LOW symbols, so the 
throughput starts to drop. When the windows are closed, the communication is possible at longer range, reaching 6 meters. 

In summary, the maximum achievable distance is more than three times better than the previous version and 6 times better than 
the original 
one. Also, the throughput is 4 times better than the previous version and more than 22 times better than the first version\cite{Wang2014vlcs}\cite{openvlc12}. 

\begin{figure}[t]
	\centering
	\includegraphics[width=0.8\columnwidth]{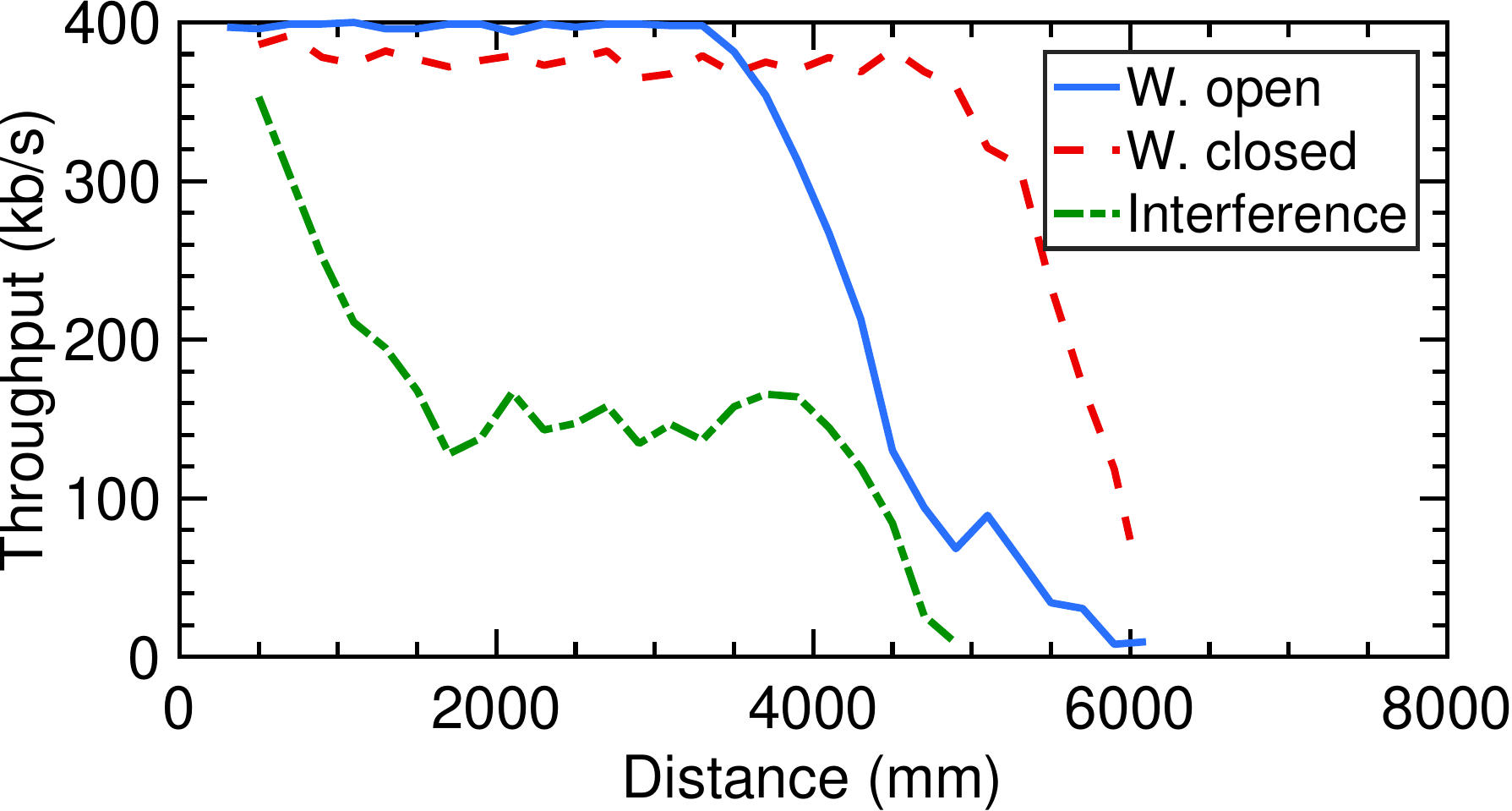}
	\caption{UDP throughput as a function of the distance.} \label{fig_th_payload}
\end{figure}

\vspace{-1ex}
\section{Limitations of the system}\label{sec_limits}

The OpenVLC1.3 platform has several \change[Ander]{limits}{limitations}, as any real system. The first one is that the throughput can not be improved without major changes in both software and hardware. Changing the ADC would increase the cost of the board and the processing power required to perform the reception. \change[Ander]{}{Also, although technically possible to use OpenVLC as a transceiver, in the current version, the throughput is maximized using one board as transmitter and one as receiver.}


In addition, the communication is UDP, as there is no return VLC link in the system. This design decision has been taken after 
realizing that:
\begin{itemize}
	\item The processing power of the BBB's PRUs is limited, and having a bidirectional system requires at least twice as processing power as an one-way link.
	\item The current trend in networking is that VLC will operate in hybrid systems, where the downlink is VLC and the uplink is 
RF\change[Ander]{, as it is not pleasant to have a light blinking for the user}{}. By using the USB interface in the BBB, users may, for instance, 
use Wi-Fi to 
send uplink data as well as ACKs for downlink VLC.
\end{itemize}

\vspace{-2ex}
\section{Conclusion} \label{sec_conclusion}
In this paper, we have presented our latest OpenVLC version and we have evaluated its performance improvements.
To the best of our knowledge, OpenVLC1.3 is the first low-cost research platform that achieves a UDP throughput of 400 kb/s using 
only low-end off-the-shelf hardware. Apart from being used for research and teaching as its predecessors, OpenVLC1.3 can enable 
real-world applications.



\bibliography{bibliography_vlc}
\bibliographystyle{IEEEtran}

\end{document}